\documentclass[conference]{IEEEtran}

\IEEEoverridecommandlockouts

\usepackage{graphicx}
\usepackage{times,amsmath,amsfonts}
\usepackage{amsmath,amssymb,amstext}
\usepackage{multirow}
\usepackage{hhline}
\usepackage{multirow}
\usepackage{cite}
\usepackage{epstopdf}
\usepackage{bm}
\usepackage{algorithm} 
\usepackage{algorithmicx}
\usepackage{algpseudocode}
\usepackage{color}
\usepackage[lofdepth,lotdepth]{subfig}
\usepackage{soul}
\usepackage{float}
\usepackage{optidef}
\usepackage{adjustbox}

\begin{document}
\title{Non-Terrestrial Network (NTN): a Novel Alternate Fractional Programming for the Downlink Channels Power Allocation }

\author{Mahfuzur Rahman, Zoheb Hassan, Jeffrey H. Reed, and Lingjia Liu 

\thanks{M. Rahman, J. H. Reed, L. Liu are with Wireless@Virginia Tech, Bradley Department of Electrical and Computer Engineering, Virginia Tech. Z. Hassan is with the Electrical and Computer Engineering Department, Universit\'e Laval, QC, Canada. 
This work was funded in part by U.S. National Science Foundation (NSF) under grant CNS-2148212.}

}

\maketitle

\begin{abstract}

Non-terrestrial network (NTN) communication has garnered considerable attention from government entities, industries, and academia in recent times. NTN networks encompass a variety of systems, including Low Earth Orbit (LEO) satellites, Medium Earth Orbit (MEO) satellites, Geostationary Earth Orbit (GEO) satellites, High Altitude Platforms (HAPS), and Low Altitude Platforms (LAPS). Furthermore, the deployment of high-throughput satellites (HTS/VHTS) in the GEO space has gained momentum. While LEO and MEO satellites offer advantages such as low latency and reduced launching costs compared to GEO satellites, this study focuses on GEO satellites due to their stationary nature and broader coverage. In traditional cellular networks, each user equipment (UE) is allocated at least one resource block (RB), which is not shared with other UEs. However, in NTN communications, where the coverage area is extensive, dedicating an RB to only one UE is an inefficient utilization of radio resources. To address this challenge, fractional programming (FP), cognitive radio, and rate splitting multiple access (RSMA) are existing technologies that enable multiple UEs to share the same RB, thereby enhancing spectral and power efficiency. This paper aims to maximize spectral efficiency, average RBG rate, and sum rate for GEO satellite systems. However, achieving this objective involves dealing with a non-convex, NP-hard problem, as it requires the logarithmic sum of different fractions. Finding a global solution to such an NP-hard problem presents significant challenges. This paper introduces a novel alternate fractional programming algorithm specifically designed to tackle these complex NP-hard problems in the context of GEO NTN cellular networks. By employing this innovative approach, the study seeks to contribute to the optimization of NTN communication systems, enabling efficient resource allocation and improved network performance.

\end{abstract}

\section{Background and Motivation}
\subsection{Introduction}
The widespread deployment of commercial 5G networks has already had a transformative impact on our lives. Researchers from both academia and industries are now actively directing their attention towards the development of the next-generation cellular technology, 6G \cite{lin, LiuAI1}. As the number of cellular data users continues to surge, meeting the demands of these users and catering to their diverse data requirements has become crucial, making the deployment of 5G and beyond imperative. The 5G wireless system is designed to support a wide range of use cases, including Enhanced Mobile Broadband (eMBB), Ultra-Reliable Low Latency Communication (uRLLC), and Massive Machine Type Communication (mMTC) \cite{LiuAI2}. In 3GPP TR Release 17, \cite{3gpp763}, 3GPP has provided guidelines on the Technical Specification for Narrow-Band Internet of Things (NB-IoT) / enhanced Machine Type Communication (eMTC) to facilitate the operation of Non-Terrestrial Networks (NTN). However, to effectively accommodate eMBB traffic everywhere, the deployment of 5G networks necessitates the implementation of 5G through the use of the NTN infrastructure.


\subsection{Non-Terrestrial Network}
The Non-Terrestrial Network (NTN) offers a significantly wider coverage footprint compared to terrestrial networks. The coverage range of non-geostationary satellites, such as Medium Earth Orbit (MEO) and Low Earth Orbit (LEO) satellites, can extend from 100 meters to 500 kilometers, while Geo-Satellites can provide even better coverage ranging from 200 kilometers to 1000 kilometers \cite{3gpp811, 3gpp821}. The expansive coverage of NTN 5G enables the provision of cellular network connectivity in areas that traditional terrestrial networks cannot reach to such an extent. Non-terrestrial 5G can be deployed in different configurations based on specific use cases. In situations where latency is not a critical factor and delay can be tolerated, Non-Terrestrial 5G can be implemented using the Bent Pipe Transparent payload, both with and without a Relay Node. On the other hand, for delay-sensitive applications requiring Ultra Low Latency Reliable Communication (uLLRC), Non-Terrestrial 5G/6G can be deployed using the Regenerative Payload, again with or without a Relay Node. In scenarios involving high-speed trains or airplanes, these vehicles can serve as the relay node for Non-Terrestrial 5G/6G deployment.





\subsection{NTN Use Cases and Motivation}
The impact of Non-Terrestrial Cellular Networks on various sectors, including general cellular usage, emergency services, disaster relief, maritime operations, US government and defense, transportation, energy, aviation, autonomous systems, and healthcare, has already begun to revolutionize our lives. Key players such as SpaceX Starlink, Amazon project Kuiper, Hughes, and Iridium are fiercely competing in the race to dominate the market. SpaceX has already launched more than 4000 Low Earth Orbit (LEO) satellites, providing global broadband connectivity, with plans to launch an additional 12000 LEO satellites in the near future. This number may further increase to a staggering 42000 LEO satellites in the coming years \cite{spacex}. Similarly, Amazon's project Kuiper aims to deploy 3236 LEO satellites to deliver fast, cost-effective, and reliable broadband services to underserved communities worldwide \cite{amazon}. Hughes, with a significant presence in North and South America, offers services through both GEO and LEO satellites. Their Jupiter series GEO satellites provide a capacity of 200 Gbps, and they have plans to launch the next-generation Jupiter GEO satellite (XXIV) in the first half of 2023, which alone will provide a capacity of 500 Gbps \cite{hughes}. Considering the wide range of applications and the guidelines provided in 3GPP Release 17 \cite{3gpp863}, and Release 16 \cite{3gpp821,3gpp811} have motivated us to explore the NTN cellular domain.

%
\section{Related Research and Contribution}
\subsection{Fractional programming}
Fractional programming (FP) is an essential branch of optimization techniques that deals with objective functions containing ratio terms \cite{frenk2005fractional}. In wireless communication systems, the data rate is often calculated using the logarithm of the signal-to-interference-plus-noise ratio (SINR), denoted as $log_2(1 + SINR)$. As wireless systems involve multiple SINR terms, solving optimization problems with these ratios requires the application of Fractional Programming (FP). However, such problems are known to be Non-Polynomial (NP)-hard, making them challenging to solve. FP techniques have proven valuable in addressing power control, energy efficiency optimization, and beamforming challenges within the communication domain \cite{Wei_Yu}.


The use of fractional programming and related optimization techniques has demonstrated their effectiveness and applicability across various domains, including wireless communication, mobile edge computing, and unmanned aerial vehicles. These techniques provide valuable tools for solving complex optimization problems and enable the efficient allocation of resources in the pursuit of improved performance and cost-effectiveness.
\subsection{Current research on NTN}
In their work \cite{Mohsen, Shuxun}, researchers provide a comprehensive overview of the current Non-Terrestrial Network (NTN) standard and highlight the challenges associated with integrating NR 5G with satellite access. In \cite{bodong_sag} the authors present detailed analysis of the space-air-ground integrated network (SAGIN), restricted and uneven allocation of computational and communication resources hinders the ability to provide reliable quality-of-service (QoS) assurances for particular types of traffic. In \cite{bodong_ag} the authors aim to reduce the overall energy usage of user equipment (UEs) through a comprehensive optimization approach. This approach encompasses optimizing user associations, adjusting uplink power levels, managing channel allocations, distributing computation capacities, and determining the three-dimensional placement of unmanned aerial vehicles (UAVs). Researchers focus on the Ka-band, which supports a maximum frequency offset of 66 kHz through the use of higher sub-carrier spacing (SCS). Higher SCS values result in shorter symbol durations, necessitating adjustments in the timing requirements for successful Random Access (RA) procedures. The performance analysis of RA procedures for NTN NB-IoT is investigated in the study conducted by the authors in \cite{carla}. They find that the standard periodicities of 40 ms and 80 ms are not suitable for RA procedures in NTN due to its large coverage footprint. Instead, an extra-large periodicity of 1280 ms is required, which leads to a reduction in the number of RA occasions. In \cite{yifei}, the authors employ Software-defined radios (SDR) B210 and Raspberry Pi4 to emulate an NTN network. They utilize a cubesat constellation for data collection pertaining to IoT services for NTN mMTC (massive Machine Type Communication). This experimental setup allows for valuable insights into the performance and behavior of NTN systems. Investigating the interference between Terrestrial Networks (TN) and Non-Terrestrial Networks (NTN), Lee et al. examine the reverse pairing scenario in their work \cite{Lee}. 



\subsection{Contribution}
In this work, we make significant contributions to the performance analysis of Non-Terrestrial Network (NTN) cellular networks, focusing on power allocation techniques. Our contributions can be summarized as follows: First, we introduce a novel alternate Fractional Programming algorithm for power allocation in the NTN downlink channels. By comparing the network performance in different scenarios, we demonstrate that our algorithm outperforms conventional approaches \cite{Wei_Yu}. Second, we extend the application of Fractional Programming and WMMSE (Weighted Minimum Mean Square Error) Power Allocation techniques, which are widely used in Terrestrial networks, to the NTN domain. This marks the first introduction of these power allocation techniques in the context of NTN networks. Third, we conduct benchmark performance analysis for both Ka-band and S-band frequencies in the NTN network. Our analysis covers both single-spot beam (SSB) and multi-spot beam (MSB) scenarios, considering various performance metrics. Specifically, we evaluate the spectral efficiency, average rate per RBG (Resource Block Group), and sum rate in different scenarios. Notably, our work represents the first benchmark performance analysis conducted for NTN networks, providing valuable insights into the network's capabilities and performance characteristics.


%
\section{System Model}
\subsection{System Overview}

\begin{figure}[htbp]
    \includegraphics[width=0.9\linewidth]{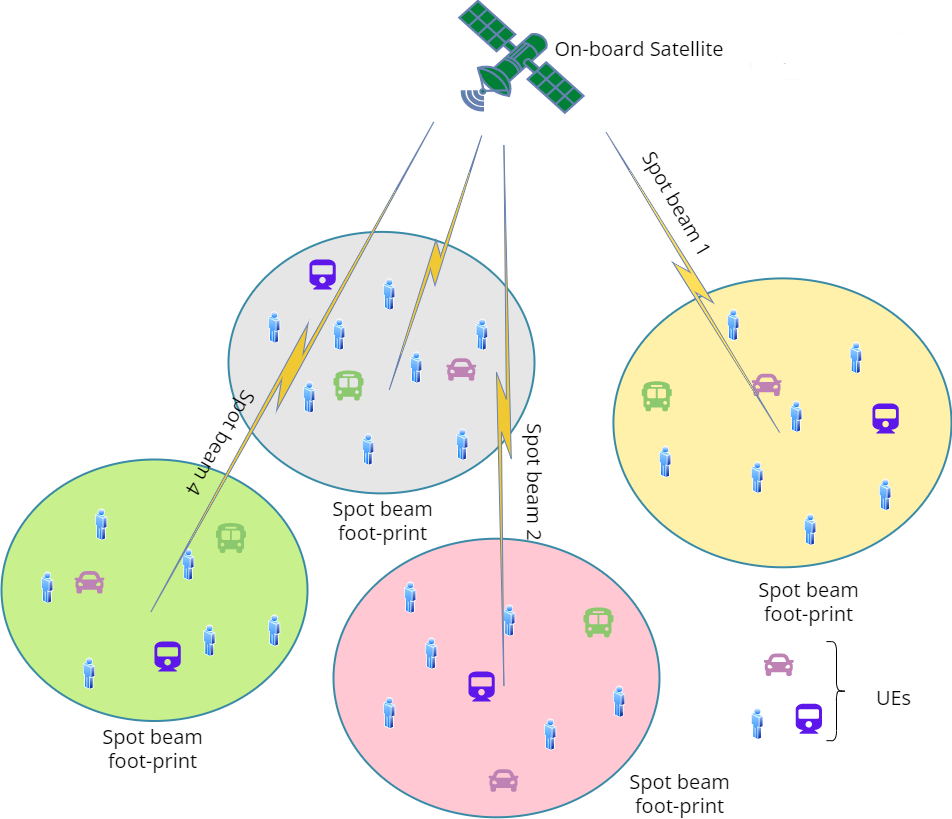}
  
\caption{Non-Terrestrial Network Architecture}
\label{system-model}
\end{figure}

We consider a GEO satellite that can operate both in single beam and multi beams operation and serves the ground UEs as shown in Fig \ref{system-model}. The service link between UEs and the Satellite is based on the 3GPP protocol stack whereas the feeder link between the satellite and the gateway is based on the satellite DVB-S2X protocol.


On-board Non-Terrestrial Network (NTN) gNB serves a set of U ground users group = $\{{ ug_1,ug_2,ug_3,....ug_k}\}$. Each ground user group can have several UEs that share the same resource block group (RBG) and each RBG consists of one or multiple resource blocks (RBs). Group of UEs share the same resource block. As specified in \cite{LiuMIMO1, LiuMIMO2}, Multi-user MIMO effectively enhances spectral efficiency however its efficacy is reduced to cell-edge users due to inter-cell interference. This traffic could be offloaded to satellite on-board gNB to enhance the overall performance.
Each of bandwidths $bw_m$, and $bw_s$ has a set of N RBs based on the 5G numerology, which is uniformly distributed into K non-overlapping RBGs to each service beam, each beam then serves each of those users' ground. The set of K RBG for the K UEs =$\{{rbg_1, rbg_2,rbg_3,.....,rbg_k}\}$, and each set of RBG consists of =$\{{rb_1, rb_2,rb_3,.....,rb_l}\}$.

Each NTN beam operates with a set of orthogonal frequencies so that inter-user group interference can be avoided. Additionally, UEs in each user group receive differentiated power levels operating on the same RBG to mitigate intra-user group interference.

\subsection{Non-terrestrial Channel model}
 
 As introduced by 3GPP \cite{3gpp811} and \cite{3gpp821}, the Non-Terrestrial channel is just the extension of the Terrestrial channel where losses associated with rain, cloud, fog, and scintillation are incorporated.

 \subsubsection{Pathloss Model}
 Pathloss of the NTN channel can be formulated as 
 
\begin{equation} \label{eqA}
PL_{total}=PL_{basic} + PL_{gas} + PL_{scin} +PL_{entry}
\end{equation}
where $PL_{total}$ is the total pathloss in dB for the radio link, $PL_{basic}$ is the basic pathloss component in dB, $PL_{gas}$ is the gaseous attenuation loss in dB, $PL_{sin}$ is the scintillation loss for the atmospheric layers in dB, $PL_{entry}$ is the associated building entry loss in dB. The basic pathloss ($PL_{basic}$ ) is associated with the signal propagation in the free space, surrounding clutter, and shadow fading effects. Gases has an effect on radio channel at frequency higher than 52 GHz, as such gases effect can be neglected. As specified in the 3GPP \cite{3gpp811} and \cite{3gpp821}, all of these scenarios are outdoor with clear sky, hence scintillation loss due the atmospheric layers, $PL_{scin}$ and building entry loss, $PL_{entry}$, both can be neglected as well. Clutter loss in basic pathloss model is significantly low, clutter is also neglect. Shadow fading expressed in dB is a zero-mean normal distribution with standard deviation 4, included in the basic pathloss equation. The free space path loss model is expressed as follows:
 
\begin{equation} 
FSPL(d,f_c) = 32.45 + 20log_{10}(f_c) +20log_{10}(d)
\end{equation}
where FSPL is the free space path loss in dB for the distance d in meter and $f_c$ is the operational frequency in GHz.

\subsection{Spectral Efficiency}
In this technique, NTN gNB splits its total available power before transmitting to its corresponding UEs. Data streams, $s_{ntn}$ are independently encoded for each $N_{ntn}$ NTN UEs. After linear precoding, data streams, $s_{ntn}$ are then transmitted over the same resource block group (RBG) by the NTN gNB. Signals transmitted from the i-th gNB are expressed as:

\begin{equation}
 {x}_{i,j} = \sum_{j=1}^J\sqrt{P_{i,j,ntn}}s_{ntn}
\end{equation}

where $P_{i,j,ntn}$ is the transmitted power from the $i$-th gNB to the $j$-th UE and $P=[P_{i,1,ntn},P_{i,2,ntn},P_{i,3,ntn},.....,P_{i,J,ntn}]$ is the power profile for all the UEs associated with the $i$-th gNB, whereas $P_{k,l,ntn}$ is the transmitted power from $k$-th TBS to the $l$-th UE and $P_{tn}=[P_{k,1,tn},P_{k,2,tn},P_{k,3,tn},.....,P_{k,L,tn}]$ is the power profile for all the UEs associated with the $k$-th TBS. At time slot $t$, received signals at $j$-th UE is expressed as:

\begin{equation}
  \text{y}_{i,j} = \sqrt{g_{i,j}^m(t)}x_{i,j} +n_0
\end{equation}

where $g_{i,j}^m(t)$ is channel gain between $i$-th gNB and the $j$-th UE during $m$-th RBG, $n_0$ is the additive white Gaussian noise  $\mathcal{N}(0,\sigma^2)$. For shake of simplicity, all the UEs have the same noise spectral density, $n_0$.
Each UE associated with the $i$-th gNB SB also receives interference from the signals transmitted for other UEs associated with the $i$-th gNB SB only. As to neglect inter-beam interference, we consider orthogonal frequencies are allocated to all gNB SBs, and UE can only connect to one SB at a time with  gNB SB. As such, rate for the non-terrestrial stream for $j$-th UE is expressed as:

\begin{equation}
\label{R_{ntn}}
R_{i,j}(m)= \log_2\left(1+\frac{P_{i,j}^{m}g_{i,j}^{m}}{\sum_{\substack{j'=1\\j'\neq j}}^{J}P_{i,j'}^{m}g_{i,j}^{m}+\sigma^2}\right)
\end{equation}


%
\section{Problem Formulation}
In this section, we would like to maximize - the overall sum rate in the Non-terrestrial cellular networks through the optimization of the transmission power allocation at the non-terrestrial base station, P; user association to the Non-Terrestrial network, $\alpha$. We consider binary optimization variable, $\alpha$ such that $\alpha$=1 if $m$-th RBG of the gNB SB is assigned to the $j$-th UE; $\alpha$=0 otherwise. The total sum rate of the Non-Terrestrial cellular network is the total achievable rate by all NTN UE for the available bandwidth. Total sum rates depend on the following two constraints.

(1) \textbf{Power constraints} $P_{i, j}^m$ is the allocated power for the $j$-th user from the $i$-th gNB SB for the duration of $m$-th RBG. $P_{i,max}$ is the maximum transmitted power of the $i$-th gNB SB for all the associated users. Then non-terrestrial power allocation constraint is: 

\begin{equation}
\label{power_allcoation}
\begin{split}
(\text{C1}) \hspace{0.2cm} & \sum_{j=1}^J  \alpha_{i,j} P_{i,j}^{(m)} \leq P_{i,max}, \forall i \in \mathcal{I}
\end{split}
\end{equation}

(2) \textbf{Cost constraints} $c_{i, j}^m$ is the cost for the $m$-th RBG of the $i$-th gNB SB. $C_{i,max}$ is the maximum allowable cost of the $i$-th gNB SB for all the associated users. Then non-terrestrial cost constraint is: 

\begin{equation}
\label{cost}
\begin{split}
(\text{C2}) \hspace{0.2cm} & \sum_{j=1}^J  \alpha_{i,j} c_{i,j}^{(m)} \leq C_{i,max}, \forall i \in \mathcal{I}
\end{split}
\end{equation}

The utility function can be designed as:
\begin{equation}
\label{Utility_func}
\begin{split}
     f(p, \alpha) = \sum_{i=1}^{I}\sum_{j=1}^{J}\alpha_{i,j}\log_2\left(1+\frac{P_{i,j}^{m}g_{i,j}^{m}}{\sum_{\substack{j'=1\\j'\neq j}}^{J}P_{i,j'}^{m}g_{i,j'}^{m}+\sigma^2}\right)\\
\end{split}
\end{equation}

The RBG of the radio resource is the cost of the cellular network. For ease of simplicity, we consider a fixed number of RGB for users' groups, hence cost constraints can be ignored here. Then we formulate a novel spectral efficiency optimization problem as follows:

\begin{equation}
\label{Opt_resource_allocation}
\begin{split}
\text{P0:} & \max_{\bm{p} \geq 0, \bm{\alpha} \in \{0,1\} } f(\bm{p},\bm{\alpha}) \\
& \text{s.t.} \quad \text{(C1)}.
\end{split}
\end{equation}
\section{Introduced Solution for NTN network}
Optimization of the spectral efficiency of \eqref{Opt_resource_allocation} involves maximization of the signal-to-interference plus noise ratio terms in \eqref{Utility_func}. After applying Lagrangian dual transform to the optimization problem \text{P0}, the equivalent objective function be: 

\begin{equation}
\label{Opt_resource_allocation1.1}
\begin{split}
\text{P1:} & \max_{\bm{p} \geq 0, \bm{\alpha} \in \{0,1\}, \bm{x_j} \geq \frac{A_j}{B_j}}  f_1(\bm{p}, \bm{\alpha},\bm{x_j}) \\
& \text{s.t.} \quad \text{(C1)}.
\end{split}
\end{equation}

Then the new introduced Lagrangian dual function can be written as - 
\begin{equation}
\label{Utility_func1.1}
\begin{split}
  &   f_{2}(p, \alpha, x_j) = \sum_{i=1}^{I}\sum_{j=1}^{J}\alpha_{i,j}\log_2\left(1+x_j\right)\\
  & -\sum_{i=1}^{I}\sum_{j=1}^{J}\alpha_{i,j}x_j + \sum_{i=1}^{I}\sum_{j=1}^{J}\frac{\alpha_{i,j}(x_j-1)A_j}{A_j-B_j},
\end{split}
\end{equation}

where $A_j=P_{i,j}^{m}g_{i,j}^{m}, B_j= \sum_{\substack{j'=1\\j'\neq j}}^{J}P_{i,j'}^{m}g_{i,j'}^{m}+\sigma^2$ and for strict duality condition case, optimum $x_j$ can be derived as:
\begin{equation}
\label{optimal_x}
    \begin{split}
      & x_j^*=\frac{P_{i,j}^{m}g_{i,j}^{m}}{\sum_{\substack{j'=1\\j'\neq j}}^{J}P_{i,j'}^{m}g_{i,j'}^{m}+\sigma^2}.  
    \end{split}
\end{equation}

After applying to quadratic transformation to \eqref{Utility_func1.1}

\begin{equation}
\label{Utility_func1.2}
\begin{split}
  &   f_3(p, \alpha, x_j,y) = \sum_{i=1}^{I}\sum_{j=1}^{J}(\alpha_{i,j}\log_2\left(1+x_j\right)\\
  & -\alpha_{i,j}x_j + 2y\sqrt{\alpha_{i,j}(x_j-1)A_j} - y^2(A_j-B_j)),
\end{split}
\end{equation}
Then applying KKT condition to eq. \eqref{Utility_func1.2} and for fixed values of $p, \alpha,x_j$, optimal $y$ can be derived when ${\partial{f_3}}/{\partial{y}} = 0 $ as: 

\begin{equation}
\label{optimal_y}
\begin{split}
y^* &=\sqrt{\frac{\alpha_{i,j}x_j}{A_j-B_j}}\\
    &=\sqrt{\frac{\alpha_{i,j}x_j}{P_{i,j}^{m}g_{i,j}^{m} - (\sum_{\substack{j'=1\\j'\neq j}}^{J}P_{i,j'}^{m}g_{i,j'}^{m}+\sigma^2)}}
\end{split}
\end{equation}

Similarly applying KKT condition to eq. \eqref{Utility_func1.2} and for fixed values of $y, \alpha,x_j$, optimal $p$ can be derived when ${\partial{f_3}}/{\partial{p}} = 0 $ as: 

\begin{equation}
\label{optimal_p}
\begin{split}
    p_{i,j}^*  &=\frac{g_{i,j}\alpha_{i,j}(x_j-1)}{y^2(g_{i,j}-\sum_{\substack{j'=1\\j'\neq j}}g_{i,j'})}
\end{split}
\end{equation}

Finally $j$-th ue's association to the $i$-th NBS can be considered as: 

\begin{equation}
\label{alpha}
\begin{split}
\alpha_{i,j}=\begin{cases}
& 1, ~ \text{if} \hspace{0.2cm} E1> 0;\\
& 0, \hspace{0.2cm} Otherwise
\end{cases}
\end{split}
\end{equation}
where $E1= \alpha_{i,j}\log_2\left(1+x_j\right)
            -\alpha_{i,j}x_j + 2y\sqrt{\alpha_{i,j}(x_j-1)A_j} - y^2(A_j-B_j)$ for $j$-th ue.
\vspace{5pt}
Then the entire process is summarized as Algorithm \ref{Algorithm1}:

\begin{algorithm}
	\caption{Transmit power allocations in  NTN network only}
	\label{Algorithm1}
	\begin{algorithmic}[]
    \State \textbf{Step 0:} Initialize $p_{i,j}$ and $x_j$ to feasible values.
    \Repeat
    \State \textbf{Step 1:} Update $y$ by \eqref{optimal_y}.
    \State \textbf{Step 2:} Update $x_j$ by \eqref{optimal_x}.
    \State \textbf{Step 3:} Update $p_{i,j}$, $\alpha$ jointly by \eqref{optimal_p} and \eqref{alpha} respectively.
    \Until the value of function $f_3$ in \eqref{Utility_func1.2} converges.
 	\end{algorithmic}
\end{algorithm}

\section{Simulation Result and analysis}

In this section, we present the numerical results and performance analysis for both single beam and multi-beam operations in the Non-Terrestrial Network (NTN). Our simulations are conducted considering a Geostationary Earth Orbit (GEO) satellite positioned at an altitude of 35,786 km, operating in either S-band or Ka-band, and serving ground User Equipment (UE) through single or multiple beams. The ground UEs are uniformly distributed within each beam, ensuring coverage across the target area.

For multi-beam operations, we employ an orthogonal frequency separation approach to minimize interference between beams. Each beam covers a designated region, equivalent to the coverage area of 19 Terrestrial Base Stations (BS). The link-level parameters are set based on the recommendations specified in \cite{3gpp811,3gpp901,3gpp821}, unless otherwise stated.

It is important to note that our analysis in this paper is focused solely on the link-level simulation. To obtain a comprehensive understanding of the NTN system, a system-level simulation is also necessary. Such a simulation should consider factors such as antenna gain calculation, high propagation delay, and large Doppler spread if applicable. In the case of Low Earth Orbit (LEO) satellites, accurate positional information, including velocity and other parameters, can be obtained through the Satellite Two Line Element (TLE) data. Therefore, incorporating TLE information becomes crucial for system-level simulations involving LEO satellites. However, for Medium Earth Orbit (MEO) and Geostationary (GEO) satellites, where their positions remain stationary with negligible velocity, TLE information may not be as critical.
 
 \subsection{Spectral efficiency of S-band single spot beam (SSB) and multi-spot beams (MSB)}

 \begin{figure}[htbp]

    \includegraphics[width=0.49\linewidth]{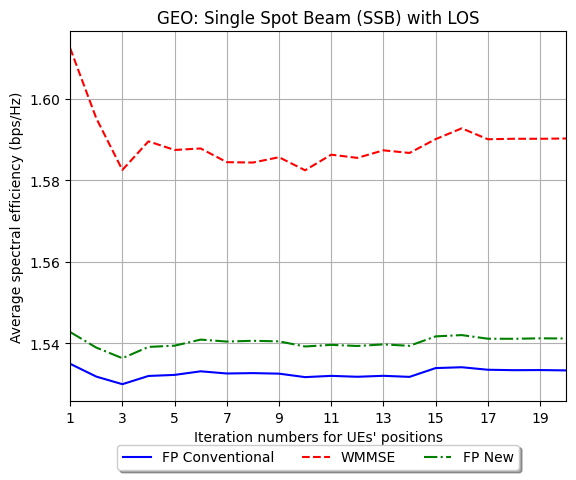}
    \hfill  
    \includegraphics[width=0.49\linewidth]{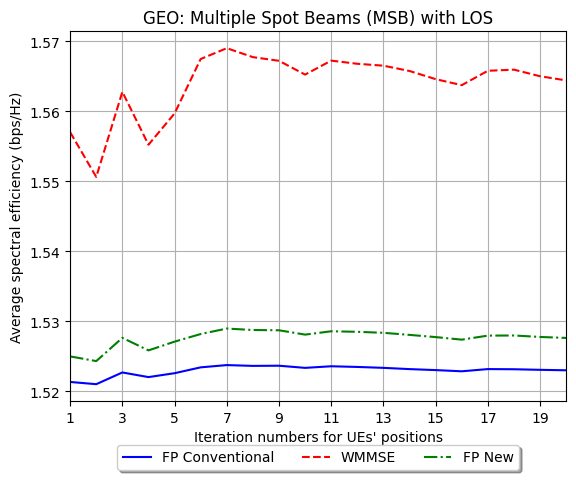}

\caption{Spectral efficiency of S-band single spot beam (SSB) and multi-spot beams (MSB)}
\label{SE_S}
\end{figure}

In this scenario, we assume that the satellite has a clear line of sight with the user equipment. The base station antenna (gNB O-RU) is mounted on the satellite, and the user equipment's height is varied from 0 to 1.5 m above the ground. Our analysis focuses on the spectral efficiency of the Non-Terrestrial Network (NTN) in the S-band (sub-6 GHz) for both single-spot beam (SSB) and multi-spot beams (MSB) 5G operations. To account for different orientations of the user equipment, we perform 20 iterations.

Figure \ref{SE_S} illustrates the spectral efficiency performance with varying user equipment positions. We observe that the performance of the NTN network is dependent on the positioning of the user equipment. In our analysis, our introduced algorithm demonstrates superior performance compared to the conventional method introduced by \cite{Wei_Yu}, for both Ka-band and S-band operations.

\subsection{Spectral efficiency of S-band and K-band}
Next, we extend our analysis to include both Ka-band (mmWave 20 GHz) and S-band (sub-6 GHz) for the Non-Terrestrial Network (NTN) 5G operation. In this case, the spectral efficiency is evaluated with respect to the variation in the number of user equipment (UEs). We conduct our analysis for a specific set of UEs in each iteration.

\begin{figure}[htbp]

    \includegraphics[width=0.49\linewidth]{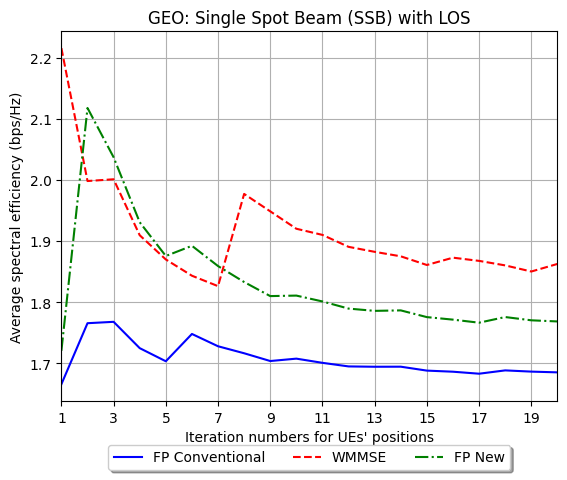}
    \hfill  
    \includegraphics[width=0.49\linewidth]{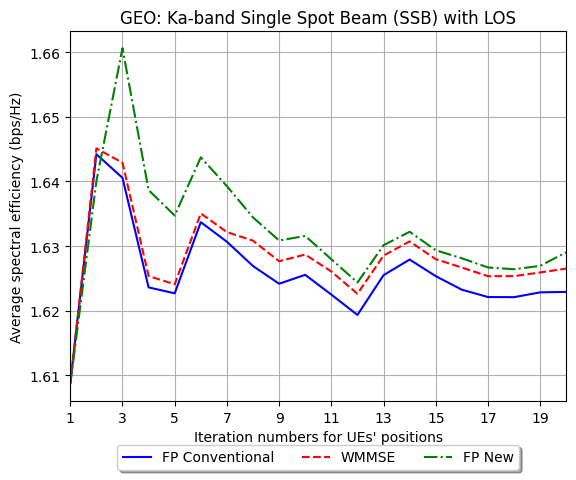}

\caption{Spectral efficiency of single spot beam (SSB) for S-band and K-band}
\label{SE_SK}
\end{figure}

Figure \ref{SE_SK} presents the results of the spectral efficiency analysis. We observe that the new introduced Fractional Programming (FP) algorithm outperforms the conventional FP approach, demonstrating higher spectral efficiency for both S-band and Ka-band operations. This improvement in spectral efficiency highlights the effectiveness of our introduced algorithm in optimizing the performance of the NTN network.

\subsection{Avg. RBG rate for S-band and K-band}
Next, we evaluate the average Resource Block Group (RBG) rate for both S-band and Ka-band operations with a single spot beam (SSB). In 5G networks, the sub-carrier spacing (SCS) is determined by the chosen numerology, which ranges from 1 to 4 [1-4]. Each RBG's capacity is influenced by the available bandwidth and the SCS. The number of RBs in each RBG is determined based on a network cost analysis.

\begin{figure}[htbp]

    \includegraphics[width=0.49\linewidth]{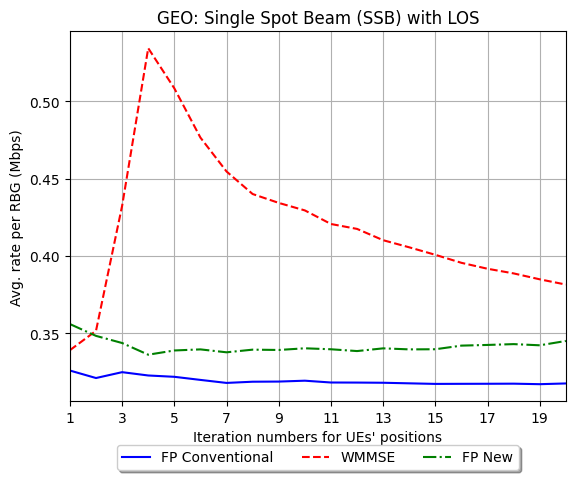}
   \hfill
    \includegraphics[width=0.49\linewidth]{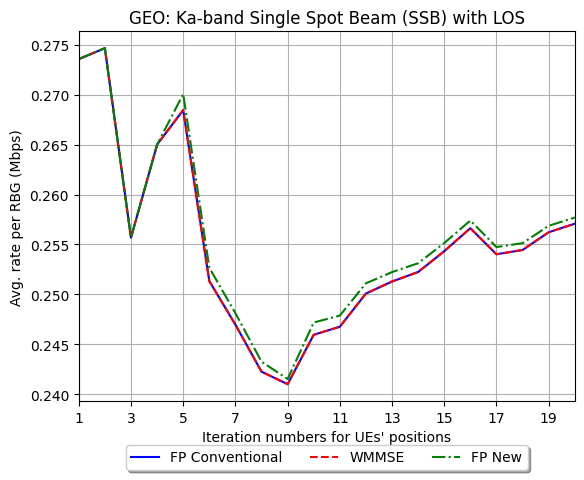}

\caption{Avg. RBG rate for S-band and K-band single spot beam (SSB)}
\label{RBG}
\end{figure}

In our analysis, we consider an SCS of 15 KHz, which is the same as LTE, and each RBG contains only one RB. Figure \ref{RBG} illustrates the results of our evaluation. We observe that the average RBG rate for all three algorithms is nearly identical for Ka-band. However, for S-band, we observe a significant improvement in the average RBG rate when using our introduced Fractional Programming (FP) algorithm compared to the conventional FP approach. This improvement highlights the superiority of our introduced algorithm in optimizing the average RBG rate for S-band operations.

\subsection{Total Sum Rate}
Finally, we evaluate the total sum rate for both S-band and Ka-band operations in both single beam and multi-beam scenarios. The total available bandwidth for S-band is 30 MHz, while for Ka-band, it is 400 MHz [1-4]. We calculate the total sum rate for S-band using the allocated 30 MHz bandwidth.

\begin{figure}[htbp]
    \includegraphics[width=0.49\linewidth]{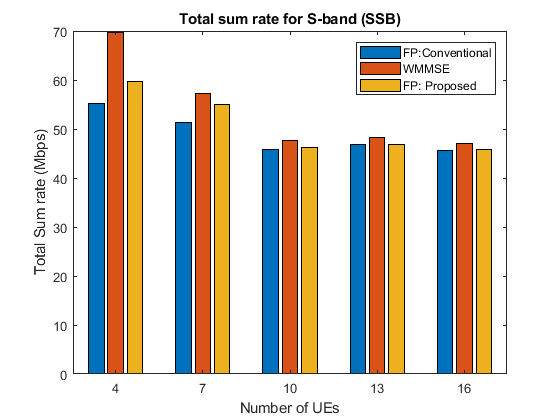}
  \hfill
     \includegraphics[width=0.49\linewidth]{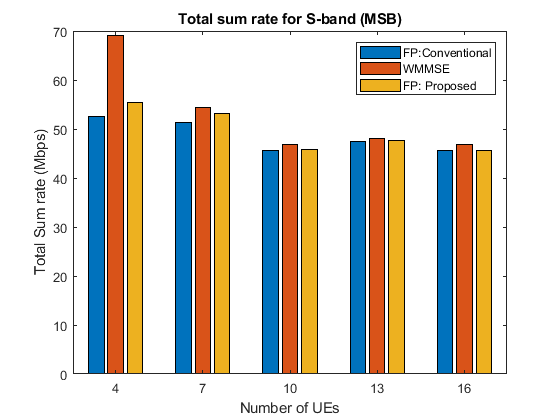}
\caption{Total sum rate for S-Band for single and multi spot beams (SSB, MSB)}
\label{sum_rate}
\end{figure}

Figure \ref{sum_rate} presents the results of our analysis. We observe that for both single spot beam (SSB) and multi-spot beams (MSB) cases, our introduced Fractional Programming (FP) algorithm outperforms the conventional FP approach introduced by Wei and Yu [5]. This improvement in total sum rate demonstrates the effectiveness of our algorithm in optimizing the overall system performance for both S-band and Ka-band operations.

\section{Conclusion}
In this study, we introduce a novel fractional programming algorithm for power allocation in both terrestrial and non-terrestrial networks (NTN). We compare the performance of our algorithm with a conventional approach in various scenarios to demonstrate its superiority. Initially, we focus on the terrestrial network and conduct a comprehensive analysis. Our introduced algorithm consistently outperforms the conventional algorithm across different scenarios. Subsequently, we extend our analysis to non-terrestrial networks. In all scenarios, our introduced fractional programming algorithm consistently outperforms the conventional algorithm when the networks are heavily crowded making it advantageous from both capital expenditure (capex) and operational expenditure (opex) perspectives.

%
%
\bibliography{IEEEabrv, main}
\end{document}